# A Data-Driven Method for Modeling Creep-Fatigue Stress-Strain Behavior Using Neural ODEs

Hao Deng*, Mark C. Messner

Applied Materials Division, Argonne National Laboratory, 9700 Cass Ave., Lemont,

IL 60439, United States of America

*Corresponding author. Email: hao.deng@anl.gov

**Abstract**

In this paper, we introduce a data-driven machine learning approach for modeling one-dimensional stress-strain behavior under cyclic loading, utilizing experimental data from the nickel-based Alloy 617. The study employs uniaxial creep-fatigue test data acquired under various loading histories and compares two distinct neural network-based ODE models. The first model, known as the black-box model, comprehensively describes the strain-stress relationship using a Neural ODE equation. To interpret this black-box model, we apply the Sparse Identification of Nonlinear Dynamical Systems (SINDy) technique, transforming the black-box model into an equation-based model using symbolic regression. The second model, the Neural flow rule model, incorporates Hooke's Law for the linear elastic component, with the nonlinear part characterized by a Neural ODE. Both models are trained with experimental data to accurately reflect the observed stress-strain behavior. We conduct a detailed comparison with the standard Chaboche model, which includes three back stresses. Our results demonstrate that the neural network-based ODE models precisely capture the experimental creep-fatigue mechanical behavior, exceeding the standard Chaboche model's accuracy. Furthermore, an interpretable model derived from the black-box neural ODE model through symbolic regression achieves accuracy comparable to the Chaboche model, enhancing its interpretability. The results highlight the potential of neural network-based ODE models to depict complex creep-fatigue behavior, eliminating the necessity for experts to define a specific, material-focused model form.

## 1. Introduction

The design of structural components subjected to cyclic loads requires an in-depth understanding of the material's behavior under these conditions. This crucial knowledge can be acquired through meticulous experimental testing and the use of reliable material models. One significant phenomenon in cyclic plasticity models, when materials are subjected to cyclic loading, is ratcheting. Ratcheting is defined as the accumulation of strain under stress-controlled cyclic loading with a non-zero mean stress. This phenomenon's description is primarily associated with kinematic hardening. Over the years, various models such as Prager, Mroz, Armstrong and Frederick, Chaboche, and Ohno and Wang have been proposed to simulate the cyclic behavior of various materials. In recent years, researchers such as Bari, Ohno, and Wang have compared these plasticity models with experimental data [1-3]. Their research has shown that Chaboche's model has been particularly effective in simulating ratcheting. As a result, the Chaboche model has gained wide acceptance and is highly regarded in the field for modeling creep-fatigue material behavior.

Where additional accuracy beyond the Chaboche model is desired, researchers often formulate material-specific, detailed constitutive models [1-3], which remains a complex and time-consuming task. It demands intricate mathematical derivations and a deep understanding of the plastic deformation mechanisms inherent in the material. However, the advent of data-driven approaches and advancements in deep neural networks have catalyzed a shift towards combining data science with solid mechanics for constitutive model development, drawing significant research interest. The burgeoning field of data science has ushered in numerous data-driven methods aimed at supplementing or supplanting traditional constitutive models. For instance, Furukawa and Yagawa [4] conceptualized inelastic material behaviors as a state-space representation, proposing a one-dimensional viscoplastic constitutive model using artificial neural networks (ANN). They utilized stress, viscoplastic strain, and internal variables as inputs, with plastic strain rate and rate of internal variable change as outputs. Furthermore, the one-dimensional mechanical response of a viscoplastic solder alloy to strain rate jumps, temperature fluctuations, or loading-unloading cycles was learned using specifically designed long short-term memory (LSTM) networks [5]. Zopf and Kaliske [6] suggested a model-free approach to characterize an uncured elastomer in three dimensions, utilizing neural networks to represent stress-stretch dependence in one dimension in conjunction with the microsphere approach. Mozaffar and colleagues [7] generated stress-strain data along different loading paths in two-dimensional space, building an elastoplastic model through recurrent neural network (RNN) learning based on this data. In many of the aforementioned studies, numerical simulations are typically employed to generate the data required to train the neural networks. This process often necessitates substantial amounts of data to yield high-quality neural network models. For instance, Mozaffar et al. [7] generated 15,000 samples with varying loading series to collect stress-strain data under different loading paths. Given that uniaxial experiments provide the most straightforward approach to acquiring stress-strain data, it is essential to develop computational models that can predict creep-fatigue responses under one-dimensional settings and complex loading paths using solely one-dimensional data. In this paper, our focus is on developing Neural ODE models to learn from this one-dimensional stress-strain data. Future work will investigate the extension of these one-dimensional models to three-dimensional models, possibly leveraging $J_2$ flow theory [8] or other established methods [9].

The relationship between neural networks and ordinary differential equations (ODEs) was quickly recognized following the introduction of residual networks by He et al. in 2016 [10]. Indeed, the similarities between the two are striking. This connection was further developed by Chen et al. in 2018 [11] with the advent of Neural ODEs, where the dynamics of the neural network are approximated using an adaptive ODE solver. Neural ODEs have demonstrated substantial potential in various fields, particularly in physical sciences (Köhler et al., 2019 [12]), and in modeling irregular time series (Rubanova et al., 2019 [13]). More recent studies have extended the application of Neural ODEs to stochastic models (Li et al., 2020 [14]). Given the promising capabilities of Neural ODEs in learning the dynamics of physical ODE systems, this paper explores the feasibility of applying Neural ODEs to learn a constitutive law describing creep-fatigue material behavior under uniaxial load.

A significant challenge in applying neural ODEs lies in efficiently calculating the gradient of the objective function with respect to the trainable parameters. Finite difference approximations of this gradient provide a straightforward yet inefficient solution. This is particularly true for complex material behavior, where the corresponding Neural Network model has a large set of trainable parameters, and for long-term cyclic loading, where each forward simulation of the material behavior is computationally costly. To overcome these challenges, calculate the model gradient using the adjoint approach, which is vital for training a neural ODE model. We have implemented an implicit Neural ODE solver in a Python package called *pyopmat* [15], which is open-source and built on the PyTorch machine learning library [16]. One of the advantages of using PyTorch as the foundation for *pyopmat* is its provision of a CUDA interface for tensors and Automatic Differentiation libraries. This means *pyopmat* can efficiently train and run models on GPUs, offering significant computational advantages.

Recently, several advanced machine learning methods have been proposed, such as operator learning. For instance, the Hierarchical Deep-Learning Neural Network (HiDeNN) has been systematically developed through the construction of structured deep neural networks (DNNs) in a hierarchical fashion. A specific variant of HiDeNN, tailored for representing the Finite Element Method (abbreviated as HiDeNN-FEM), has been established [17]. HiDeNN-FEM has been further adapted for various applications, ranging from diverse Partial Differential Equations (PDEs) to Ordinary Differential Equations (ODE) solvers. Liu et al. have also developed a nonlinear version of HiDeNN-FEM [18]. Additionally, the Convolution Hierarchical Deep-Learning Neural Network Tensor Decomposition (C-HiDeNN-TD) has been introduced for high-resolution topology optimization [19]. Moreover, Sebastian et al. have developed neural operators specifically for Bayesian inverse problems [20].

Symbolic Regression (SR) is a type of supervised learning where models are formed from analytic expressions. This approach is often tackled using a multi-objective optimization framework that aims to minimize both the prediction error and the complexity of the model. Unlike traditional methods that adjust specific parameters within a complex, predefined model, SR explores a variety of straightforward analytic expressions to find models that are both accurate and easy to understand. Historically, scientists have engaged in a form of manual SR, using their intuition and a trial-and-error process to identify simple, yet precise, empirical formulas. These formulas could then pave the way for significant theoretical advancements, like the discoveries that led to classical and quantum mechanics. SR algorithms modernize this process, leveraging advanced computing to evaluate a broader array of expressions than could be managed by intuition alone. The initial use of SR as a scientific tool date back to the 1970s and 1980s, including the development of the Bacon tool and its successors like Fahrenheit [21, 22]. These tools used heuristic methods to exhaustively search through potential expression trees, successfully uncovering various basic physics laws from simplified data. Additionally, the adoption of genetic algorithms has enhanced SR, offering a more adaptable search space and reducing the need for initial assumptions about the expressions, further advancing the field's capabilities [23].

The need for Symbolic Regression arises as machine learning models, notably large ones like transformers, grows increasingly complex with billions of parameters to enhance prediction accuracy. However, the complexity of these models often leads to a lack of clarity about their functioning and their limitations. This is particularly problematic in fields like physical sciences, where understanding the causal relationships and underlying principles is crucial. SR stands out as a promising approach to address these challenges by offering models that are not only accurate but also interpretable and insightful. In scientific research, the value of a model often lies not just in its accuracy but in its ability to provide a clear, concise representation of the underlying phenomena. This is where SR shines, as it seeks to discover equations that strike a balance between simplicity and accuracy. Such equations may not always be the most precise, but they offer insights and understanding that are invaluable, especially in hypothesis-driven areas like the physical sciences. While machine learning models like Transformers and LSTMs may excel in accuracy, they often fall short in offering insights into the problem's nature, making them less reliable and predictable in certain contexts. Standard physical models, in contrast, are favored for their ability to encapsulate physical phenomena through simplified equations, providing predictability and reliability even if they are not the most accurate. This is where the Neural Ordinary Differential Equation (ODE) model comes into play, distinguishing itself from other time series machine learning models through its interpretability and explanatory power. By grounding the model in real physical systems, it offers insights into how each term or internal variable interacts, aiding in the deeper understanding of system dynamics. This can empower physicists and scientists to develop new theoretical models, making SR an invaluable tool in the progression of machine learning and scientific discovery.

The primary contribution of this paper is the innovative introduction of Neural ODE models to learn the constitutive laws based on experimental data, coupled with the use of symbolic regression to interpret the fitted Neural ODE model.. Most existing literature focuses on using numerical simulation data as ground truth, with few studies applying machine learning to actual experimental creep-fatigue data. Moreover, a distinct advantage of the Neural ODE model is its ability to learn the constitutive law based on small datasets, which is beneficial given the high cost of obtaining experimental data. The rest of the paper is structured as follows: Section 2 presents the formulation of the standard Chaboche model and the Neural ODE models. Section 3 offers a detailed comparison of the training results between the Chaboche and Neural ODE models based on experimental stress-strain data. Section 4 uses the symbolic regression technique to interpret the Neural ODE model, generating an equation-based representation of the physical model, which makes the black-box model more explainable. Finally, Section 5 concludes the paper and outlines directions for future research.

## 2. Chaboche and Neural ODE Approaches for Constitutive Law

In this section, we introduce the standard Chaboche model along with two Neural ODE models, each featuring a distinct mathematical formulation. These two models are referred to as the 'Black-box model' and the 'Neural flow rule model'.

### 2.1 Standard Chaboche Model

We begin with the standard Chaboche model, presenting the equations in a one-dimensional setting. To extend the model to three dimensions simply requires the replacement of the viscoplastic rate equations with their $J_2$ flow counterparts, alongside updating the backstress evolution equation to utilize symmetric rank two tensors instead of signed scalars. The model begins from Hooke's law:

$$\dot{\sigma} = E(\dot{\varepsilon} - \dot{\varepsilon}_{vp}) \tag{1}$$

where $E$ is the Young's modulus, $\dot{\varepsilon}$ is the total mechanical strain rate, here taken from the experimental strain history, and $\dot{\varepsilon}_{vp}$ is the viscoplastic strain rate. This model inherently accommodates plastic-creep, eliminating the need for explicit coupling of creep and plastic strain models. The model evolves the isotropic and kinematic hardening as following,

$$\dot{\varepsilon}_{vp} = \langle \frac{|\sigma - X| - \sigma_0 - K}{\eta} \rangle^n \text{sign}(\sigma - X) \tag{2}$$

where $\sigma$ is the stress, $X$ is the total kinematic hardening, $\sigma_0$ a threshold stress, $K$ the isotropic hardening, $\eta$ the constant flow resistance, and $n$ the rate sensitivity. In general, the kinematic hardening $X$ can be decomposed into multiple backstresses, for this work we use three:

$$X = \sum_{i=1}^{3} X_i \tag{3}$$

The evolution of each backstress is governed according to the standard Chaboche model, as follows,

$$\dot{X}_i = \frac{2}{3} C_i \dot{\varepsilon}_{vp} - \gamma_i X_i |\dot{\varepsilon}_{vp}| \tag{4}$$

$C_i$ and $\gamma_i$ represent the kinematic hardening and dynamic recovery parameters, respectively. Additionally, the model incorporates Voce isotropic hardening as follows,

$$\dot{K} = \delta(R - K)|\dot{\varepsilon}_{vp}| \tag{5}$$

In Equation (5), $R$ represents the saturated value of the isotropic hardening variable, indicating the maximum limit that $K$ can reach under plastic deformation. The aforementioned Chaboche model forms a system of ordinary differential equations. Given appropriate initial conditions, this system can be solved using implicit methods. To calibrate the Chaboche model based on experimental data, we employ the L-BFGS algorithm [24] to identify the model's parameters. The loss function to minimize for the calibration problem can be formulated as follows:

$$loss = \sum_{j=1}^{n_{exp}} \sum_{i=1}^{n_{time}} \left(\sigma_{j,i}^{model} - \sigma_{j,i}^{exp}\right)^2 / (n_{exp} \cdot n_{time}) \tag{6}$$

In the above formulation, $n_{exp}$ represents the number of experiments, and $n_{time}$ refers to the number of time steps. The outer sum encapsulates each experiment in the database, while the inner loop spans each individual measurement (time, strain, stress, etc.) in the experimental record. This model is configured for strain control, where the forward model evaluation provides the stress given the strain, temperature, and time.

## 2.2 Neural ODE Model: Black box Model

The Chaboche model utilizes ODEs to depict the time-dependent behavior of creep-fatigue materials. Broadly, in lieu of employing physics-based material models, we could leverage machine learning models to learn the dynamic, time-dependent material behavior, treating the strain-stress rate relationship as a black box. Historically, differential equations have been instrumental in describing a system's behavior by detailing its instantaneous dynamics. These equations have often been derived from theoretical frameworks, such as Newtonian mechanics, Maxwell's equations, or epidemiological models for infectious diseases, with parameters inferred from observations. As these equations typically do not offer closed-form solutions, numerical approximations are necessary. Recently, the use of ordinary differential equations (ODEs), parameterized by millions of learned parameters (termed 'neural ODEs'), has gained prominence. These are calibrated for latent time series models, density models, or as substitutes for deep neural networks. This suggests the potential for discovering equivalent creep-fatigue constitutive models that can be empirically observed using neural ODEs. A natural approach is to directly learn a parametric system of controlled ODEs

$$\frac{d\boldsymbol{x}(t)}{dt} = f(\boldsymbol{x}(t), \theta, u) \tag{7}$$

Here, $\boldsymbol{x}(t)$ represents a vector corresponding to the stress state and some arbitrary set of internal variables, while $f$ denotes a neural network characterized by a specific network architecture. The external input, such as strain or strain rate, is represented by $u$. $\theta$ represents the learnable parameters within the Neural Network. This algorithm aims to decipher the behavioral model of an unknown material using a neural ordinary differential equation (Neural ODE). It starts by defining a set of state variables represented by stress (σ), strain (u), and internal variables (h). The process involves feeding these variables into the Neural ODE, which computes the rate of change in stress and internal variables over time. The basic implementation is presented as follows, ($NN$ denotes a neural network)

**Algorithm 1** Neural ODE

1: Define state variables $\mathbf{x} = [\sigma, \mathbf{h}, u]^T$, where $\sigma$ is stress, $u$ is strain, and $\mathbf{h}$ represents internal variables. The dimension of $\mathbf{x}$ is $n + 2$, where $n$ is the dimension of $\mathbf{h}$.
2: Formulate the neural ODE for the constitutive model:

$$\frac{d\bar{\mathbf{x}}}{dt} = \text{NN}(\mathbf{x})$$

3: Here, $\bar{\mathbf{x}} = [\sigma, \mathbf{h}]^T$ and $\frac{d\bar{\mathbf{x}}}{dt}$ represents $[\frac{d\sigma}{dt}, \frac{d\mathbf{h}}{dt}]^T$. Note that only $\sigma$ (stress) is observable.
4: The neural ODE can be explicitly written as:

$$\frac{d[\sigma, \mathbf{h}]^T}{dt} = \text{NN}([\sigma, \mathbf{h}, u]^T)$$

5: In this setup, the input to the Neural Network (NN) is $n + 2$ dimensional, including stress, strain, and internal variables. The output of the NN is $n + 1$ dimensional, including stress and internal variables.

---

This model can be trained either by backpropagating directly through an ODE solver or by implicitly differentiating through the ODE solutions using an adjoint method, as suggested by Chen et al. (2018) [11]. Given the potential for neural ODEs to be stiff, we utilize an implicit solver implemented in the open-source framework *pyoptmat* [25], with the computations accelerated by GPUs.

We primarily consider cases where function $'f'$ is formed by a Residual Network (ResNet) [26]. ResNet, or Residual Network, is chosen due to its ability to efficiently deal with vanishing gradient problems that arise in deep networks. It does so by using skip or shortcut connections, also known as residual connections, which allow gradients to be back-propagated to earlier layers, making the training process of deep architectures more effective and stable. A ResNet is a type of deep learning model where the weight layers learn residual functions relative to their respective inputs. It is unique in its use of skip connections, or "residual connections", which perform identity mappings that are then added to layer outputs. This architecture allows ResNets to support deep learning models with scores tens or even hundreds of layers, thereby simplifying training and improving accuracy.

The mathematical equation for a simple ResNet block, often referred to as a "cascade network", can be represented as follows. Given an input $x$, the output $y$ of a ResNet block is calculated as:

$$y = F(x, W) + P \cdot x \tag{8}$$

Here, $F(x, W)$ represents the residual mapping to be learned. In the case of a simple ResNet, this could be two layers with a ReLU activation function in between. $W$ represents the weights of these layers. The operation $F(x, W) + P \cdot x$ is performed elementwise, which is possible because the dimension of $P \cdot x$ is the same as the dimension of $F(x, W)$. This equation indicates learned mapping $F(x, W)$ is not mapping from $x$ to $y$ directly, but rather it maps from $x$ to a modified version of $P \cdot x$ and then adds this to the output. This is a variant of the original Residual Network concept, where instead of adding the original input $x$ to the output of the residual mapping $F(x, W)$, we are adding a scaled version of the input $x$. This could be interpreted as a form of weighted shortcut connection, where the input $x$ is scaled by a factor $P$ before being added to $F(x, W)$. Additionally, when dealing with multiple inputs and outputs, the scaling factor $P$ can indeed be a matrix rather than a scalar.

Thus, we can formulate the Black box model, which learns the stress-strain relationship for creep-fatigue material behavior using a ResNet, as follows:

$$\frac{d\sigma(t)}{dt} = NN(\sigma(t), h, \varepsilon) + P \cdot \varepsilon \tag{9}$$

In this equation, $\sigma$ represents stress, $\varepsilon$ stands for strain, $h$ is the internal variable, and $NN$ symbolizes a fully connected neural network. $P$ represents an additional parameter that can also be learned from the training data.

### 2.3 Neural ODE Model: Neural Flow Rule Model

To complement the prior black-box model, we propose a novel Neural ODE-based model, referred to as the Neural Flow Rule Model. This model is defined by the following formulation:

$$\frac{d\sigma(t)}{dt} = E \cdot \left(\frac{d\varepsilon(t)}{dt} - dh/dt\right) \tag{10}$$

where $h$ represents the history evolution. We hypothesize that the rate of change of this history, $dh/dt$, can be modeled as a residual network (ResNet) as follows:

$$\frac{dh}{dt} = R\left(\sigma(t), h, \frac{d\varepsilon(t)}{dt}\right) \tag{11}$$

In this equation, instead of directly employing a network to depict the stress-strain relationship, we incorporate Hooke's law, representative of the elastic part of the deformation, into the model. Here, the network, denoted as $R$, represents the nonlinear portion of the relationship and needs to be trained based on experimental data. $(d\varepsilon(t))/dt$ signifies the total strain rate, while $h$ stands for some arbitrary set of internal variables. This model's primary advantage is that it incorporates the well-proven concept of an inelastic flow surface, utilizing a neural network to model only nonlinear inelastic strain and the internal state. As such, it can be viewed as a hybrid model that integrates some elements of a physical model.

This modeling approach will be less flexible compared to a black box neural ODE model, but also has several advantages. First, it incorporates some basic physical knowledge about the behavior of the system by mimicking a classical plastic flow rule. This helps in reducing the parameter space and making the model more interpretable. Secondly, it can better generalize to scenarios beyond the training data range as its response is constrained to follow a familiar plastic flow response for all inputs. Equation (10) can be explained as follows,

- $E \cdot \frac{d\varepsilon(t)}{dt}$: This term refers to the linear elasticity described by Hooke's Law, which states that the strain in a material is proportional to the applied stress. $E$ is the modulus of elasticity (or Young's Modulus) which is a material property describing the material's stiffness. This term takes into account the initial linear elastic response of the material when stress is applied.
- $R\left(\sigma(t), h, \frac{d\varepsilon(t)}{dt}\right)$: This term is the ResNet (residual neural network) model, which aims to represent the nonlinear, inelastic flow portion of the stress-strain relationship. The ResNet model is trained on experimental data and is designed to capture the nonlinear, time-dependent behavior exhibited by creep-fatigue materials. The inputs to this model are stress $\sigma(t)$, internal variables $h$, and strain rate $\frac{d\varepsilon(t)}{dt}$. Here, $\sigma(t)$ represents the current stress state, $h$ are internal variables which may include things like the material's previous loading history or microstructural state, and $d\varepsilon(t)/dt$ is the rate of change of strain. It is a function (represented by a ResNet) of stress, internal variables, and strain rate. This non-linear term captures the complexities of material behavior, particularly the inelastic or irreversible deformations.

- $E \cdot \left(\frac{d\varepsilon(t)}{dt} - R\left(\sigma(t), h, \frac{d\varepsilon(t)}{dt}\right)\right)$: The entire equation represents the stress rate as a function of the difference between the strain rate and the predicted value from the ResNet model, scaled by the elastic modulus. This implies that the rate of stress change in the material is governed by the discrepancy between its elastic response and the complex, time-dependent deformation behavior (as captured by the ResNet model). The elastic part of the model tries to restore the material to its original state, while the ResNet part captures the remaining nonlinear, irreversible deformation behavior. In this sense, the equation can be interpreted as a balance between the elastic recovery forces and the inelastic, or plastic, deformation forces caused by creep and fatigue. The material's response (rate of change of stress) is governed by its elastic behavior modulated by its inelastic behavior (as learned by the ResNet from data).

This algorithm is designed to understand the behavior of a material whose properties are not fully known. It starts by defining a vector called $x$, which includes stress (denoted as $\sigma$) and some internal variables (denoted as $h$). The strain rate, represented by input $u$ or $\dot{\varepsilon}$, is controlled during the process. The key part of this algorithm involves a Neural Ordinary Differential Equation (Neural ODE) to model how the material behaves under different loading conditions. This Neural ODE calculates stress and the internal variables over time. The basic implementation is demonstrated as follows,

---
**Algorithm 1** Neural Rule
---
1: Define state variable vector $\mathbf{x} = [\sigma, \mathbf{h}]^T$, where $\sigma$ is stress, $\mathbf{h}$ are internal variables, and $u$ is the strain rate $\dot{\varepsilon}$ under strain-controlled conditions. Thus, the dimension of $\mathbf{x}$ is $n + 1$, where $n$ is the dimension of $\mathbf{h}$.

2: Formulate the Neural ODE for identifying the constitutive model as:

$$\frac{d\sigma}{dt} = E \cdot u - [\text{NN}([\sigma, \mathbf{h}, u]^T)]_1 = E \cdot \dot{\varepsilon} - [\text{NN}([\sigma, \mathbf{h}, \dot{\varepsilon}]^T)]_1$$

$$\frac{d\mathbf{h}}{dt} = [\text{NN}([\sigma, \mathbf{h}, \dot{\varepsilon}]^T)]_{2:\text{end}}$$

3: Note: Only the first term $\sigma$ is observable. The above term can be more explicitly written as:

$$\frac{d}{dt}[\sigma, \mathbf{h}]^T = \begin{bmatrix} E \cdot \dot{\varepsilon} - [\text{NN}([\sigma, \mathbf{h}, \dot{\varepsilon}]^T)]_1 \\ [\text{NN}([\sigma, \mathbf{h}, \dot{\varepsilon}]^T)]_{2:\text{end}} \end{bmatrix}$$

4: Minimize the sum of squared differences in the Neural Network training process:

$$\text{Minimize} : \sum (\sigma - \sigma_{\text{exp}})^2$$

where $\sigma_{\text{exp}}$ is the experimental stress result, the real measurement.

---

Where the subscript $[\cdot]_1$ denotes the first term of the vector, $[\cdot]_{2:end}$ represents the terms from the second to the end, and T indicates the transpose of the matrix. When compared to the Chaboche model, the Chaboche model employs kinematic hardening to describe cyclic plasticity, adding backstresses to represent translations of the material flow surface. This model has a long history successfully representing the general evolution of stress-strain hysteresis in high temperature cyclic behavior, but often fails to capture complex details of the material response without additional modifications. The Neural Flow Rule Model, in contrast, learns material behavior (including potentially more intricate behaviors) from the data directly, offering a more adaptive approach. Yet, by retaining a term from Hooke's Law, it ensures that the foundational principle of material elasticity is not lost, merging adaptability with basic mechanics.

Mechanically, the integration of a neural network into the equation acknowledges that materials, especially under cyclic loads, exhibit behaviors beyond simple elasticity. These behaviors, influenced by microstructural changes, loading history, and other factors, can be complex and multifaceted. A neural network, especially a ResNet with its capability to learn long-range dependencies and avoid vanishing gradient problems, offers a mathematically solution to capture these intricacies. In conclusion, the Neural Flow Rule Model serves as a bridge, mathematically and mechanically, between traditional constitutive models and the adaptability of neural networks.

## 2.4 Training Process and Model Initialization

### 2.4.1 Solving Stiff Neural ODEs Using the Open-source Package *pyoptmat*

Modeling constitutive models for creep fatigue using Neural ODEs essentially involves reconstructing stiff ODEs based on limited experimental data. Stiff systems of ordinary differential equations (ODEs), coupled with sparse training data, are common challenges in scientific problems. The algorithm we employ is derived from the open-source code *pyoptmat*, developed by our research group. This code efficiently integrates stiff ODE systems through time and calculates parameter gradients using the adjoint method, employing implicit, vectorized methods.

Stiff ODEs are, somewhat ambiguously, defined as systems that are challenging to integrate using explicit numerical time integration methods. Effectively solving implicit stiff ODEs is crucial for accurately identifying unknown stiff ODEs. As outlined in the paper [1], the major contributions of *pyoptmat* lie in addressing two distinct challenges:

- Parallelization of Numerical Time Integration: This strategy involves vectorization across both the number of time series and a chunk of time integration steps. Applicable to both implicit and explicit integration schemes, it is particularly effective for implicit time integration. The method achieves speedups, including both forward and backward passes, of over 100 times for data-sparse problems compared to sequential algorithms that batch only over the number of independent time series.
- Efficient GPU Implementation: This implementation encompasses implicit time integration schemes with performance levels sufficient for training ODE and neural-ODE models using conventional gradient-descent algorithms. By vectorizing both over the number of independent time series and a batch or "chunk" of sequential time steps, we effectively assemble the implicit system of ODEs. The block-bidiagonal structure of the linearized implicit system, particularly for the backward Euler method, allows for further vectorization using parallel cyclic reduction (PCR). This dual-axis vectorization supplies a broader bandwidth of computations to the computing device, fully utilizing modern GPUs even for comparatively data-sparse problems. Speedups exceeding 100 times relative to standard sequential time integration are achieved. These acceleration benefits, alongside the advantages of implicit, vectorized time integration, are demonstrated through various example problems, including both analytical stiff and non-stiff ODE models, as well as neural ODE models. Further details regarding this acceleration can be found in [1].

Here we use the open-source code *pyoptmat* to solve the neural ode inverse problem for constitutive modeling, the detailed numerical implementation can be found in Ref [1].

### 2.4.2 Model Parameters Initialization for Neural ODE

Effective initialization of Neural ODE models is crucial for achieving successful model performance. Recent developments in neural network initialization offer various strategies, among which He initialization [27] has gained popularity. However, initializing the network at a single point often introduces random factors that can affect the effectiveness of this process. To enhance robustness, we integrate the Monte Carlo method with He initialization. This approach involves initializing the network after sampling '$n$' times, where '$n$' is an integer representing the number of different initial points. In this study, we chose $n = 30$ for all examples.

This process generates multiple initial points, after which we employ *pyoptma*' to solve the Neural ODE and compute the Mean Squared Error (MSE) for each initial point. Subsequently, the initial point yielding the minimum MSE is selected for an optimized 'good start'. Such a method enhances the robustness of the initialization, thereby contributing to more effective training of the Neural ODE model.

2.5 Effective 1D Representation of 3D J2 Plasticity in Neural ODE Framework

To simplify the complexity of 3D plasticity modeling while retaining essential material behavior, we employ a Neural ODE framework that captures $J_2$ plasticity through effective (scalar) measures of stress and strain. In $J_2$ plasticity theory, yielding under multiaxial loading conditions depends on the second invariant of the deviatoric stress tensor, $J_2$. For a given stress tensor $\sigma_{ij}$, the deviatoric stress tensor $s_{ij}$ is defined as:

$$s_{ij} = \sigma_{ij} - \frac{1}{3}\delta_{ij}\sigma_{kk} \quad (12)$$

where $\sigma_{kk}$ is the trace of the stress tensor, and $\delta_{ij}$ is the Kronecker delta. The yield condition is then defined by the effective (or von Mises) stress $\sigma_{eff}$, given as:

$$\sigma_{eff} = \sqrt{\frac{3}{2}s_{ij}s^{ij}} \quad (13)$$

The material yields when $\sigma_{eff}$ reaches the yield stress $\sigma_y$, allowing us to express yielding and plastic flow in terms of this scalar effective stress, regardless of the specific multiaxial stress state. Once yielding initiates, plastic deformation is governed by an associated flow rule, where the plastic strain rate tensor $\varepsilon_{ij}^p$ is proportional to the deviatoric stress. To simplify plasticity to a 1Dscalar form, we define the effective plastic strain rate $\dot{\varepsilon}_p$, capturing cumulative plastic deformation across all tensor components:

$$\dot{\varepsilon}_p = \sqrt{\frac{2}{3}\dot{\varepsilon}_{ij}{}^p \dot{\varepsilon}^{pij}} \quad (14)$$

This scalar quantity $\dot{\varepsilon}_p$ provides a uniaxial equivalent of the 3D plastic deformation, compressing the multiaxial behavior into an effective 1D form. Using these effective quantities, we construct a Neural ODE model that evolves $\sigma_{eff}$ and $\varepsilon_p$ over time under complex loading conditions. The Neural ODE is formulated as:

$$\frac{d}{dt}\begin{bmatrix}\sigma_{eff}\\ \varepsilon_p\\ \chi\end{bmatrix} = f\left(\begin{bmatrix}\sigma_{eff}\\ \varepsilon_p\\ \chi\end{bmatrix}, \theta, u\right) \quad (15)$$

where $f$ is a neural network function parameterized by $\theta$, designed to learn the time-dependent evolution of $\sigma_{eff}$ and $\varepsilon_p$ under loading history $u$. Here, $\chi$ represents additional internal variables that capture material-specific behaviors, such as back stress for kinematic hardening, isotropic hardening variables, or damage indicators that evolve with cyclic loading. These internal variables allow the model to incorporate a broader range of material responses beyond simple plastic flow. This approach allows the Neural ODE to capture complex behaviors such as cyclic hardening or softening, creep, and stress relaxation without solving full 3D tensorial equations.

This approach also provides a means for extending the trained 1D model back to three dimensions. The neural model provides a 1D map between the effective stress and the effective strain rate. If we assume the material follows a $J_2$ flow rule we can extend the 1D model back to 3D by replacing the 1D evolution of stress in Algorithm 1 with the 3D rule:

$$\dot{\sigma}_{ij} = C_{ijkl}\left(\dot{\varepsilon}_{kl} - NN_1 \frac{s_{kl}}{\sqrt{s_{mn}s_{mn}}}\right)$$

where the neural flow rule $NN_1$ as well as the neural hardening rule $NN_2$ are evaluated with the scalar effective strain and effective stress.

Non-uniaxial experimental data would be required to validate this extension of the model to three dimensions, unfortunately such data is seldom available. However, this assumption would allow using the model in, for example, three dimensional simulations of component behavior.

## 3. Experimental Data for Alloy 617

### 3.1 Experimental Setup and Data Preparation

To assess the efficacy of machine learning methodologies with test data, we examine a dataset of creep-fatigue trials on Alloy 617, supplied by the Idaho National Laboratory. Creep-fatigue tests typically cycle uniaxial specimens of the material under strain control, maintaining constant strain at either the maximum tension end of the cycle, the maximum compression end, or both for some period of time. The test data were obtained from uniaxial samples of Alloy 617 tested at 950°C under both monotonic and cyclic loading conditions. Figure 1 illustrates a typical test setup where the sample strain, monitored by an extensometer, controls the servo-hydraulic load frame to apply strain-controlled loading. The stress experienced by the sample is measured using a load cell. Thermocouples regulate the sample temperature through a furnace. For uniaxial experiments, testing typically continues until failure, whereas for cyclic experiments, it proceeds until there is a 20% reduction in load from the peak flow stress. The Fig. 1 depicts a uniaxial, axisymmetric test sample mounted in the machine, connected to thermocouples for temperature control, and attached to an extensometer for strain measurement in the gauge section. While the furnace door is shown open in the picture, it will be closed during the test to maintain the required temperature.

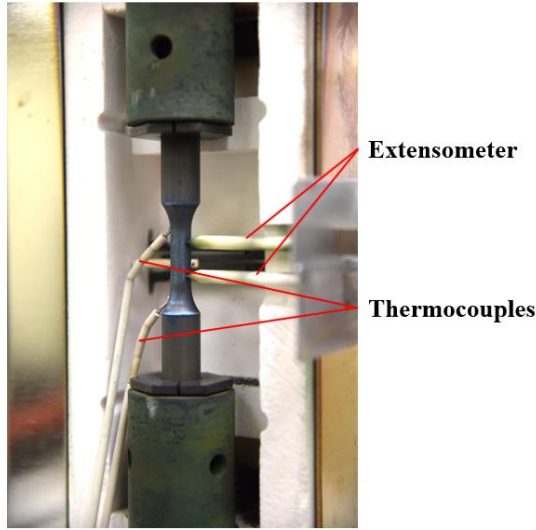

Figure 1. An Experimental Test Setup for Uniaxial Testing of Alloy 617 at 950°C

The dataset at our disposal covers a variety of scenarios, including fully-reversed, strain-controlled cyclic load spanning 0.3%, 0.6%, and 1.0% strain ranges along with hold times of 0, 3, 10, and 30 minutes. Notionally, the tests were conducted at 950°C. For each trial, we have access to the complete experimental record, which notes the gauge stress, strain, and temperature as functions of time.

This data has previously been utilized for model calibration for Alloy 617 (as per Messner and Sham, 2021 [28]). We are repeating the calibration process here for comparison between the conventional Chaboche model and the Neural ODE model. Our calibration process utilizes only a portion of the complete dataset outlined in Messner and Sham (2021), opting for 19 out of the original 33 creep-fatigue datasets due to data quality considerations. The experimental results utilized for model calibration are depicted in Fig. 2. For the purposes of training the stress-strain curves have been normalized within the range of [-0.5, 0.5].

The normalization equation is given by:

$$x' = \frac{(x - min(x))}{(max(x) - min(x))} - x_0 \qquad (12)$$

In this equation, $x'$ is the normalized value of $x$. $x_0$ is an offset to ensure that the initial state is 0. This process scales the values of $x$ between -0.5 and 0.5, which can be beneficial for many machine learning algorithms.

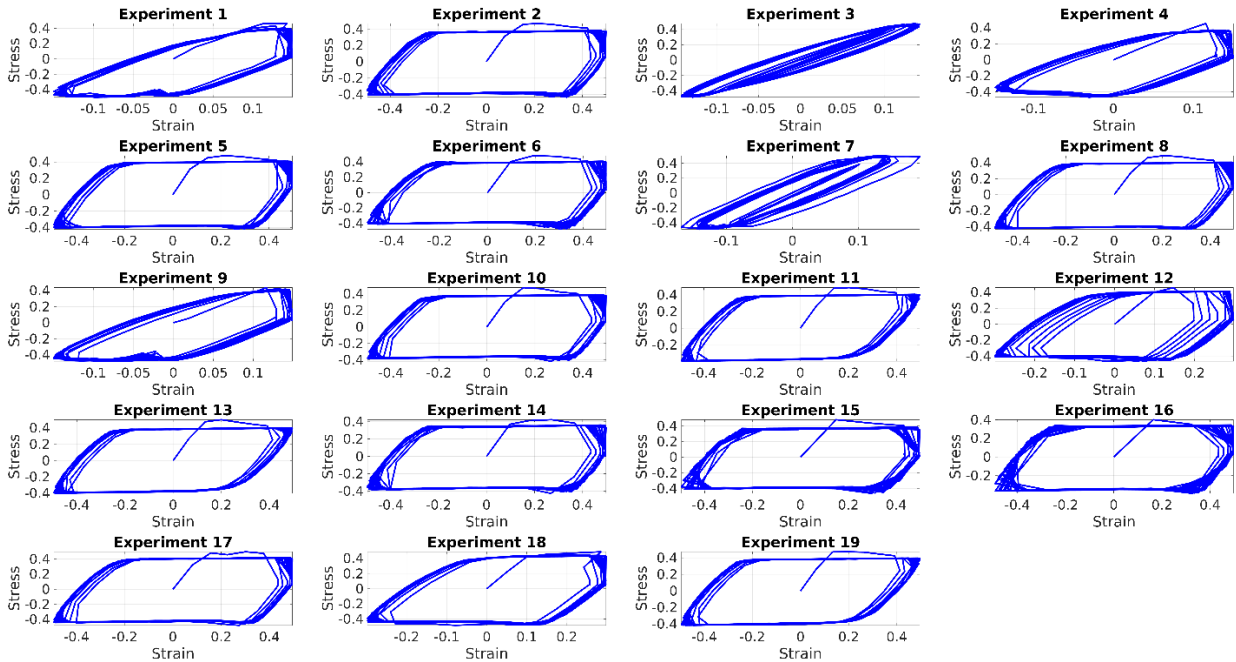

Figure 2. Creep-fatigue experimental result after normalization

## 3.2 Characteristics of cyclic loads

This data set includes creep-fatigue tests describing the influence of dwell time on the evolution of the material flow stress and the stress-strain hysteresis. Fig. 3 presents the strain's logarithmic time evolution under strain-controlled cyclic conditions, while Fig. 4 illustrates the corresponding stress response. To enhance clarity, we chose to present data on only 4 out of the 19 experiments in separate figures, as illustrated in Fig. 5 and Fig. 6. Upon examination of the stress-strain curves, several distinct behavioral patterns are found based on Fig. 2-6:

- Linear Elastic Behavior: In the initial segments of the curves, a linear relationship between stress and strain is evident. This represents the elastic behavior of the material, where deformation is temporary and the material returns to its original shape upon load removal.
- Plastic Deformation: Beyond the elastic limit, the curves begin to deviate from linearity, signaling the onset of plastic deformation.
- Fatigue: The cyclic nature of the curves showcases the fatigue behavior. As the material undergoes repeated loading and unloading, microscopic cracks may initiate and propagate, leading to eventual material failure. Before failure, the development of fatigue damage manifests as cyclic softening. The variability in the amplitude and frequency of these cycles across experiments provides insights into different fatigue scenarios and their impact on material longevity.
- Creep Indicators: The tests include holds at constant strain, which provide information on the material creep behavior. Creep, a time-dependent deformation, occurs due to prolonged exposure to stress. Over time, even below the yield strength, materials can slowly and continuously deform, a behavior evident in several experiments.

In analyzing the stress-strain curves, a notable overshoot is observed in the first cycle of loading, where the peak stress surpasses that of subsequent cycles. This behavior could stem from several factors: initial microstructural realignments offering an initially higher flow resistance, inherent material hysteresis, or viscoelastic properties leading to time-dependent variations. Additionally, external influences like test machine dynamics or variations in initial load application could contribute to this initial peak. Overall, the dataset provides a comprehensive depiction of the intricate behaviors of materials subjected to cyclic loads. However, additionally, the data, derived from real experiments, might contain errors. These inaccuracies could arise from inherent material characteristics, manufacturing processes, or experimental conditions: factors not present in data from theoretical model, which are typically devoid of noise and errors. Nevertheless, the ability of a data-driven model to accurately reflect real experiments is crucial. This distinction forms a significant point of comparison between this study and previous literature.

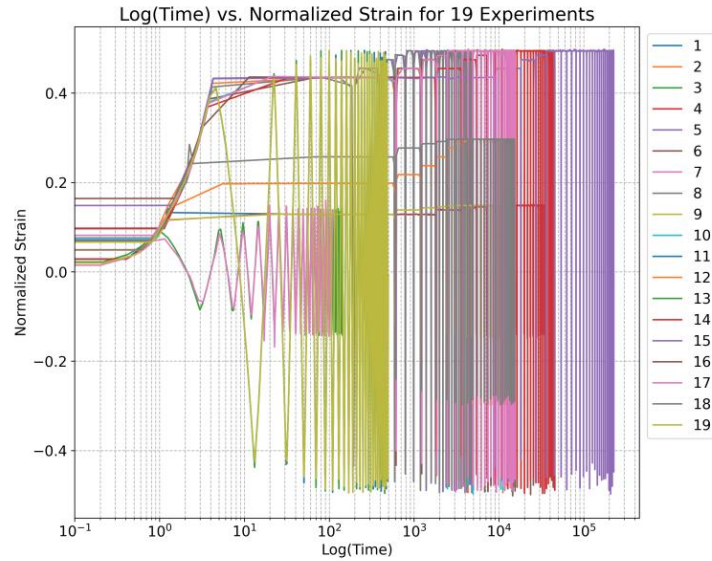

Figure 3. Logarithmic time evolution of the normalized strain in strain-controlled cyclic experiments

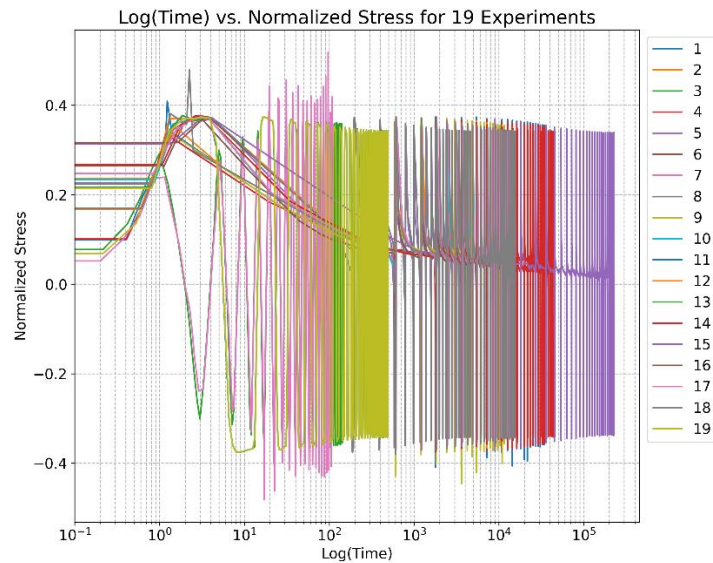

Figure 4. Logarithmic time evolution of normalized stress response in strain-controlled cyclic experiments

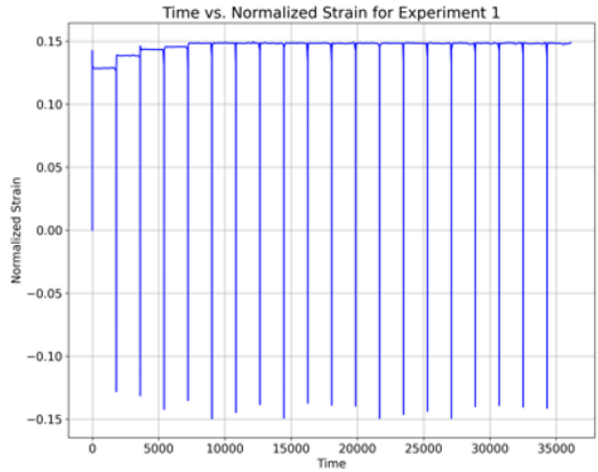
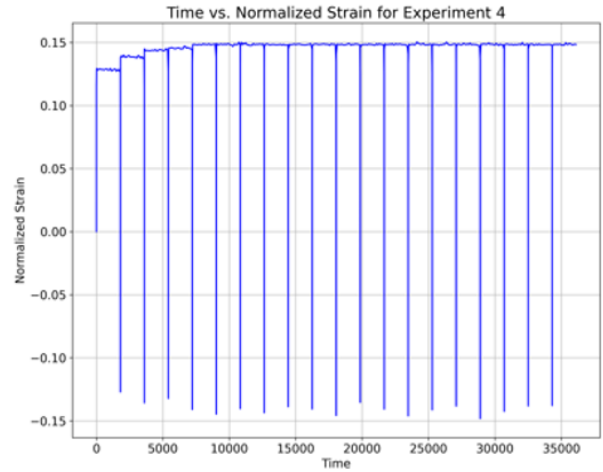
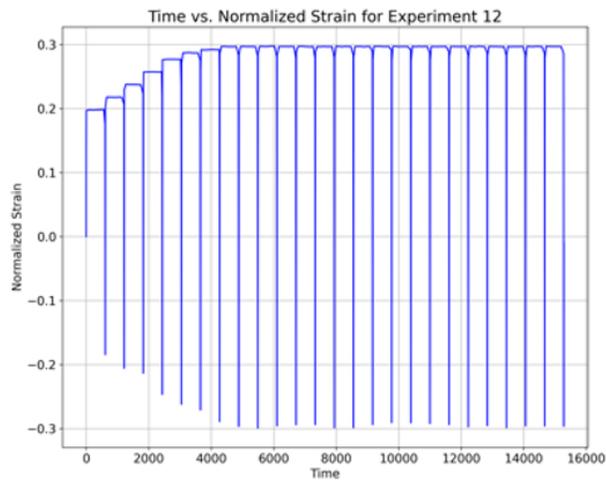
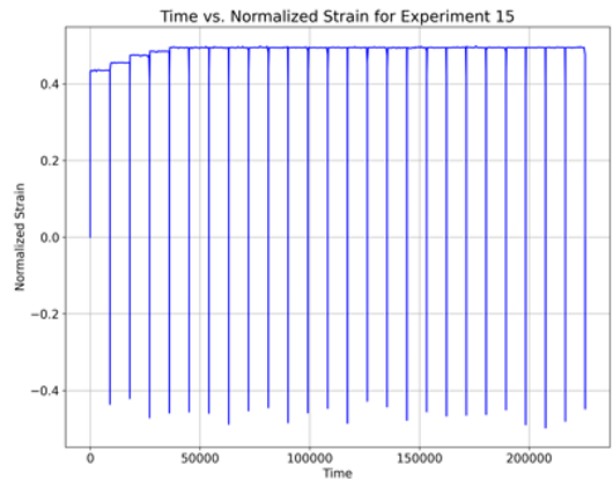

Figure 5. Time evolution of strain in strain-controlled cyclic experiments

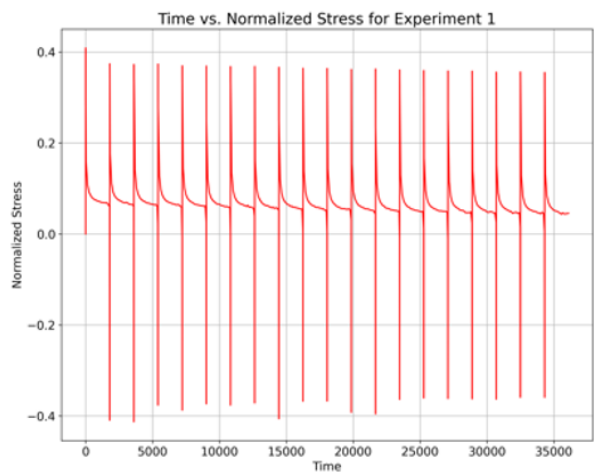
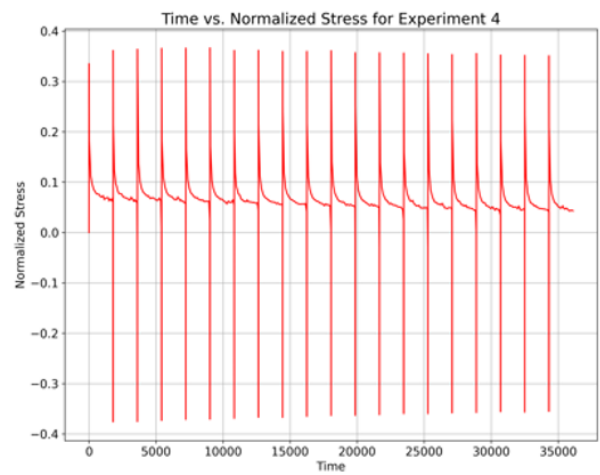

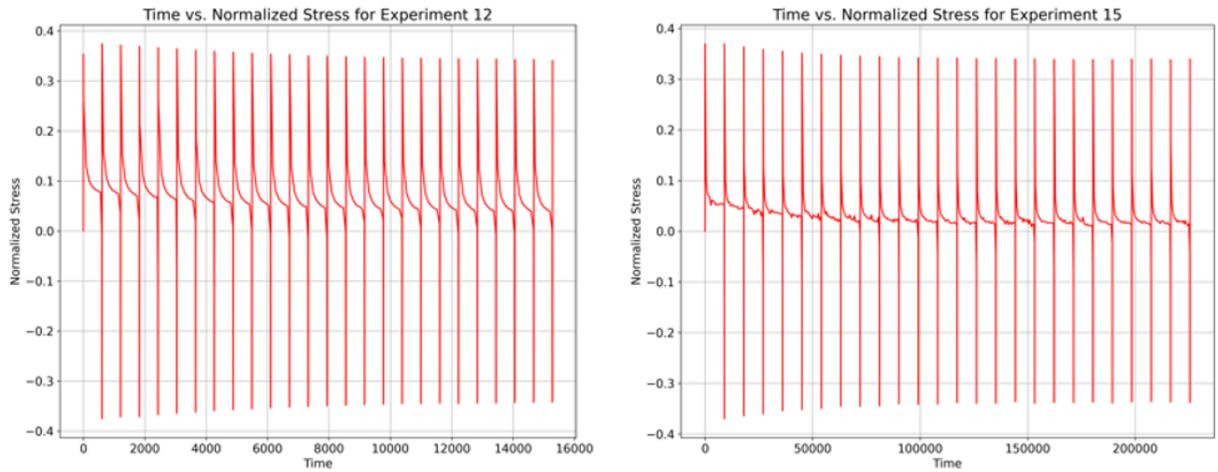

Figure 6. Time evolution of stress in strain-controlled cyclic experiments

The figures in this section provide a comprehensive overview of the experimental data used for training and validating the model. In total, 19 creep-fatigue tests on Alloy 617 were conducted, covering strain ranges of 0.3%, 0.6%, and 1.0%, along with hold times of 0, 3, 10, and 30 minutes, at a test temperature of 950°C. Figure 2 shows the normalized stress-strain responses across these 19 experiments, capturing diverse behaviors in cyclic plasticity and fatigue damage. Figure 3 illustrates the logarithmic time evolution of normalized strain, while Figure 4 presents the corresponding normalized stress responses, both under strain-controlled cyclic conditions. To provide further clarity, Figures 5 and 6 display the time evolution of strain and stress, respectively, for selected experiments, illustrating detailed cyclic responses over time. The initial gradual increase in the applied strain range, as seen in Figure 5, indicates a "ramp-up" phase, where the strain is progressively applied to avoid a sudden load that could cause material instability or undesired transients at the start of the experiment. This approach helps in stabilizing the material response and reduces the risk of introducing abrupt stresses that could affect the accuracy of the cyclic testing data. Another reason is the experimental setup or control system, where the strain amplitude is incrementally adjusted until the target strain range is reached. This gradual approach can ensure a more controlled loading application, allowing the material to adapt before entering the steady cyclic loading phase.

These figures highlight the range of loading conditions and cyclic behaviors included in the training dataset, emphasizing the diverse material responses that the model needs to learn for effective generalization. However, due to the high cost and complexity of conducting additional experiments, the dataset is limited to these 19 experiments.

Table 1 summarizes the key experimental conditions and analysis metrics for strain-controlled cyclic tests conducted at various strain ranges and hold times. For each experiment, parameters such as mean and maximum hold times, strain statistics (mean, max, min), total duration, and the mean strain rate.

Table 1. Experimental Conditions and Analysis Metrics for Strain-Controlled Cyclic Tests

| Experiment_Label | Mean_Hold_Time (s) | Max_Hold_Time (s) | Mean_Strain (%) | Max_Strain (%) | Min_Strain (%) | End_Time (s) | Mean_Strain_Rate (%/s) |
|---|---|---|---|---|---|---|---|
| Experiment_1 | 109.7436 | 110.7477 | 0.1309 | 0.2988 | -0.2992 | 36099.86 | 0.114 |
| Experiment_2 | 152.5901 | 152.5901 | 0.1294 | 0.9968 | -0.9894 | 15494.4147 | 0.164 |
| Experiment_3 | 0.2526 | 0.2605 | 0.0008 | 0.285 | -0.2845 | 143.7729 | 0.1854 |
| Experiment_4 | 104.6484 | 113.9539 | 0.1388 | 0.3007 | -0.2965 | 36109.861 | 0.1003 |
| Experiment_5 | 72.0961 | 72.0961 | 0.0806 | 0.9945 | -0.9886 | 4994.43 | 0.1745 |
| Experiment_6 | 192.3615 | 192.5901 | 0.3057 | 0.9914 | -1.0031 | 45494.4084 | 0.1288 |
| Experiment_7 | 0.1956 | 0.1956 | 0.0053 | 0.3864 | -0.3369 | 108.69 | 0.2995 |
| Experiment_8 | 137.0486 | 137.3632 | 0.0443 | 0.9902 | -0.9883 | 4994.9427 | 0.1804 |
| Experiment_9 | 76.2658 | 76.7557 | 0.0648 | 0.298 | -0.299 | 12099.87 | 0.1703 |
| Experiment_10 | 151.6296 | 151.7891 | 0.1289 | 0.9933 | -0.9946 | 15494.4075 | 0.1645 |
| Experiment_11 | 0 | 0 | -0.0003 | 0.9912 | -0.987 | 494.8994 | 0.1925 |
| Experiment_12 | 95.3758 | 96.4219 | 0.1202 | 0.5949 | -0.5984 | 15287.142 | 0.1521 |
| Experiment_13 | 0.6668 | 0.6668 | 0.0007 | 0.9912 | -0.9919 | 494.8877 | 0.1933 |
| Experiment_14 | 191.35 | 192.991 | 0.3083 | 0.9964 | -0.9926 | 45494.3904 | 0.1271 |
| Experiment_15 | 551.6551 | 587.0237 | 0.5396 | 0.9969 | -0.9952 | 225494.4102 | 0.0805 |
| Experiment_16 | 33.8401 | 33.8478 | 0.5783 | 0.9944 | -0.9854 | 15494.442 | 0.0726 |
| Experiment_17 | 149.8172 | 152.0294 | 0.1305 | 0.9905 | -0.9869 | 15494.944 | 0.1631 |
| Experiment_18 | 98.9003 | 99.2724 | 0.1216 | 0.5955 | -0.5954 | 15296.64 | 0.1528 |
| Experiment_19 | 0 | 0 | -0.0001 | 0.9911 | -0.9902 | 494.8936 | 0.1931 |

## 4. Results and Discussion

First, we calibrate the Chaboche model using the experimental data. The L-BFGS optimization method [27] is employed to calibrate the 12 parameters. The calibrated parameters for the model at a temperature of 950°C are presented in the following Table 1.

Then, to establish a comparative analysis, we utilize the black-box model to fit the stress-strain response. After testing various network structures, including different layer counts, neuron quantities, and activation functions, we discovered that the best performance was achieved using a ResNet with 3 layers, each layer consisting of 256 neurons and leveraging GeLU (Gaussian Error Linear Unit [29]) as the activation function. We explored various neural network architectures through a trial-and-error approach to optimize model performance. Configurations with 1 to 5 layers and 16, 32, 64, 128, 256, or 512 neurons per layer were tested, along with different activation functions, including GeLU, ReLU, and Sigmoid. After comparing the fitting and validation results, we found that a network with 3 layers, each containing 256 neurons and a GeLU activation function, provided the best performance. ResNet was chosen over a feedforward neural network due to its faster convergence rate and numerical stability during optimization.

Appropriate weight initialization in a ResNet is critical for successfully fitting a model. Inappropriate initialization can cause issues such as slow convergence, suboptimal solutions, or the phenomena of vanishing/exploding gradients. In the literature, various initialization methods have been proposed. A popular choice for ResNets is the He initialization [27], named after Kaiming He, the first author of the ResNet paper. This method is specifically designed for ReLU (or variants) activation functions, which are often used in ResNets. His scheme initializes the weights in the network with values from a Gaussian distribution with mean 0 and variance $2/n$, where $n$ is the number of input units. This strategy prevents vanishing or exploding gradients by maintaining the variance of the outputs of each layer approximately equal to the variance of its inputs. In this study, He initialization was applied to achieve better weight initialization. Additionally, we used He initialization multiple times to create various initial points, and we selected the point with the lowest training loss as the real initial point. Specifically, we used He initialization 30 times to generate an initial point dataset. Concurrently, we split the experimental dataset into a training set (15 stress-strain curves) and a validation set (4 stress-strain curves). The Adams method [30], with a learning rate of 1e-3, was used to optimize the neural networks.

Table 2: Calibrated Parameters for Chaboche Model

| Parameter | Value | Unit |
|---|---|---|
| Young's modulus ($E$) | 135738.5 | MPa |
| Rate sensitivity ($n$) | 4.72 | - |
| Drag stress ($\eta$) | 617.5 | MPa |
| Threshold stress ($\sigma_0$) | 0.0 | MPa |
| Isotropic softening stress ($R$) | -9.9855 | MPa |
| Isotropic softening rate ($\delta$) | 11.00 | - |
| Backstress 1 hardening rate ($C_1$) | 100.0 | MPa |
| Backstress 2 hardening rate ($C_2$) | 144.27 | MPa |
| Backstress 3 softening rate ($C_3$) | 10.0 | MPa |
| Backstress 1 dynamic recovery ($\gamma_1$) | 5.15 | - |
| Backstress 2 dynamic recovery ($\gamma_2$) | -31.65 | - |
| Backstress 3 dynamic recovery ($\gamma_3$) | 0.98 | - |

Figure 7 illustrates the training history for the black box model, which is displayed on a logarithmic scale. After 8000 iterations, both training and validation losses have converged. Some fluctuation is noticeable at the beginning of optimization; however, the overall trend of the convergence history is smooth.

Figures 8 and 9 compare the black box model predictions to the experimental data, for a portion of the training and validation sets. The blue solid line represents the experimental data, while the red dashed line indicates predictions from the black box model. For comparison, predictions from the standard Chaboche model for the validation dataset are presented in Fig. 10, with the green line representing Chaboche predictions and the blue line showing the experimental data. Both the standard Chaboche model, using a three back-stress formulation, and the black box model deliver good results as evident from these figures. However, while both models provide credible predictions for the validation dataset, the black box model achieves a lower Mean Squared Error (MSE), indicating its superior predictive capability in comparison to the standard Chaboche model. Furthermore, during the elastic response and initiation of the plastic deformation stage, the black box model exhibits a relatively better fit, especially in capturing the overshoot phenomenon (as shown in Fig. 8 with green dashed line) observed during the first loading cycle. This is a phenomenon the Chaboche model consistently fails to capture and in fact cannot capture with the standard model form. This suggests that the black box model offers enhanced flexibility to capture detailed material behavior, compared to the Chaboche model.

While the black box model results are comparatively better than the Chaboche model, discrepancies remain between the experimental data and predictions. Such inconsistencies could arise from measurement noise or potential local minima challenges faced during the training process. Despite numerous attempts with varying neural network architectures and training algorithms, obtaining an exact match between the black-box Neural ODE model and actual experimental data was elusive. Several factors could contribute to this difficulty.

First, as a continuous machine learning model, a Neural ODE may struggle to capture noise or experimental errors. The second factor is that, in comparison with methods like LSTM [7], the Neural ODE model includes latent physics assumptions that limit its flexibility in fitting any given data, including measurement errors. However, these constraints also offer certain advantages. The Neural ODE model is resistant to noise, is not easily affected by measurement uncertainties, and can reveal underlying physical phenomena or laws. Furthermore, compared to existing LSTM black box models for time-series prediction, the Neural ODE model requires a smaller dataset. In this case, just 15 training datasets and performs well on 4 validation datasets.

From these examples, we can conclude that the Neural ODE model achieves better fitting accuracy as the physics-based Chaboche model, indicating that the black-box Neural ODE models can effectively capture creep-fatigue material behavior even without any underlying physical concepts. Consequently, Neural ODE models could potentially be used for modeling the behavior of new materials, such as composite materials. Figure 14 provides a detailed comparison of the Mean Squared Error (MSE) between the Neural ODE black box model and the Chaboche model.

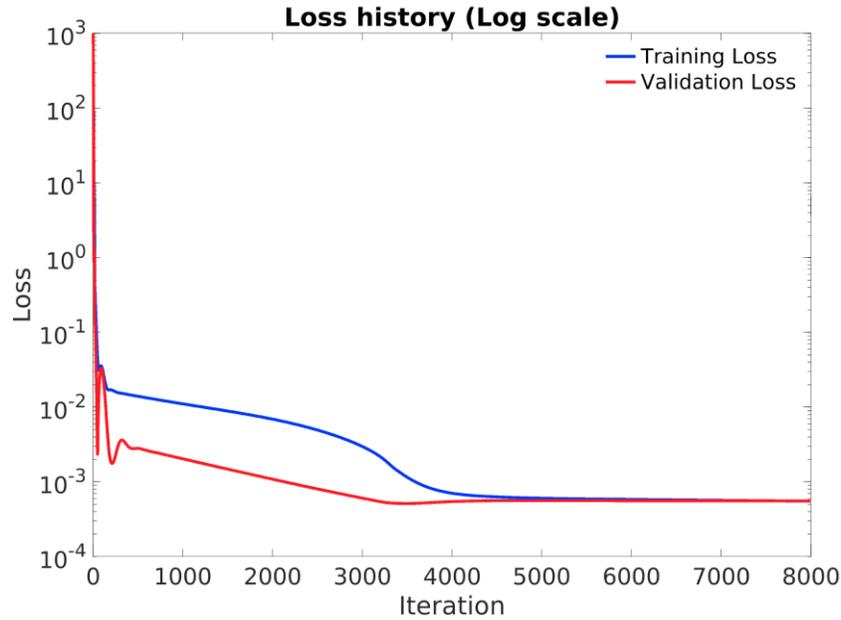

Fig. 7. Loss history for black-box

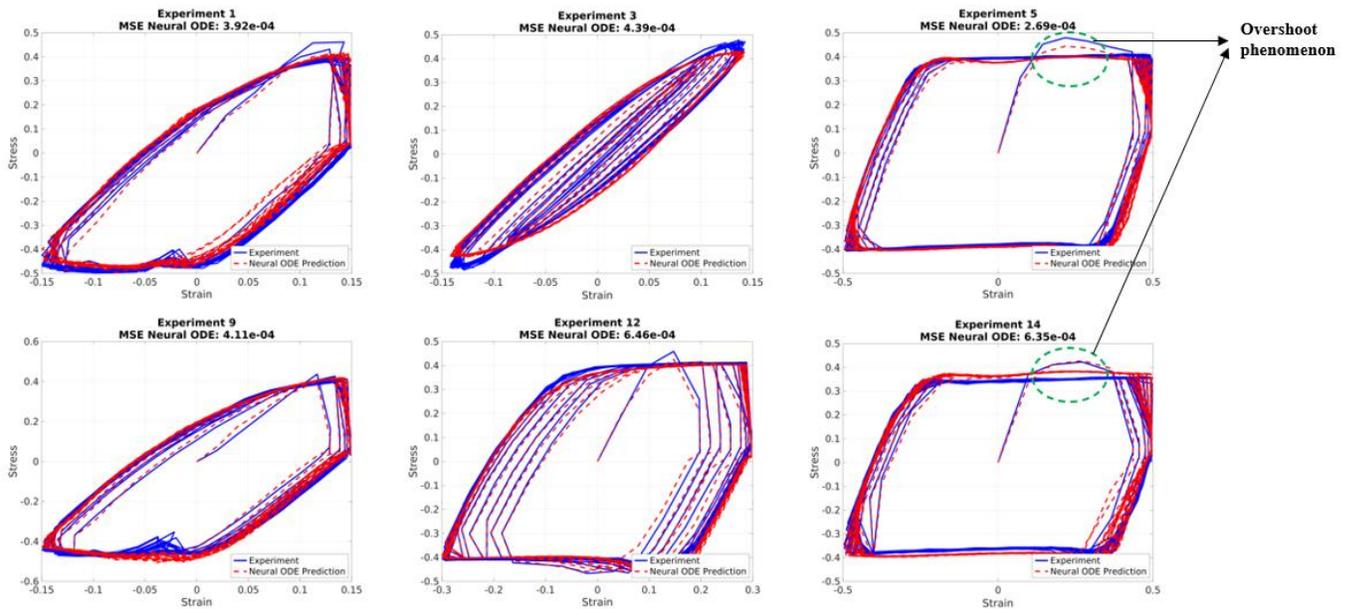

Fig. 8. Partial Fitting Results from the Black-Box Model for the Training Dataset.

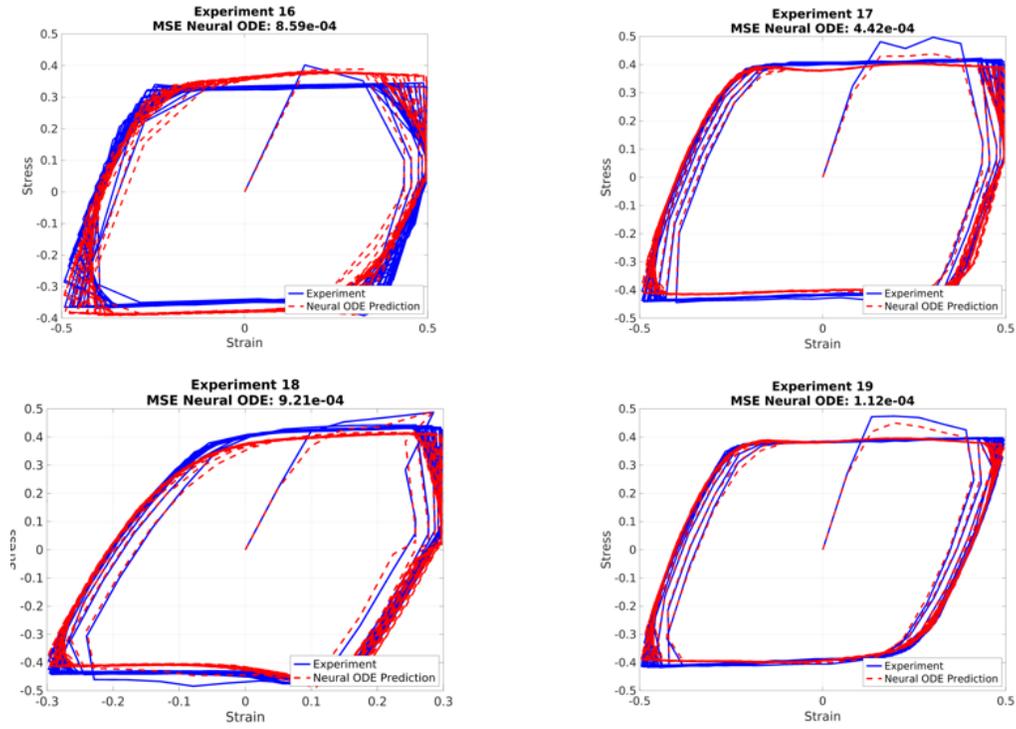

Fig. 9. The Black-Box Model for validation dataset

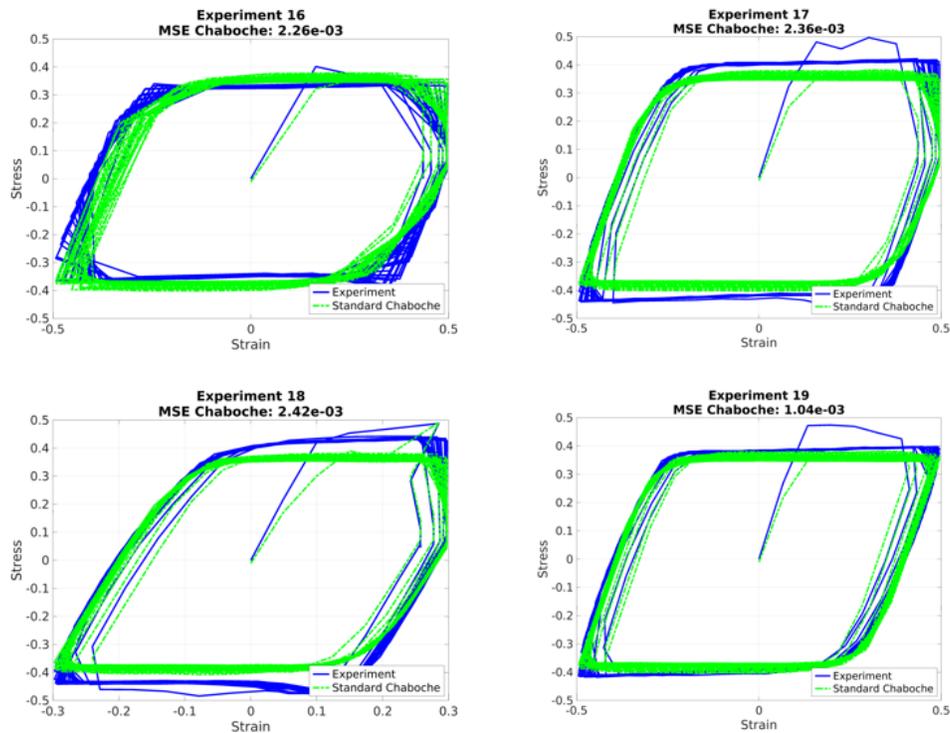

Fig. 10. The Standard Chaboche Model for validation dataset

For the Neural flow rule model, we adopt the same network structure as the black-box model. We employ the Adam optimization algorithm with a learning rate of 1e-3. The $L_1$ norm is utilized for regularization to mitigate overfitting. To monitor convergence, we use an early stopping criterion based on the validation loss. In line with our previous strategy, He initialization is employed to set the initial network parameters. The convergence history is depicted in Figure 11, where the red solid line represents the validation loss, and the blue solid line signifies the training loss. The loss function is presented on a logarithmic scale. As is evident from Figure 11, the training process reaches convergence after approximately 8000 iterations. The overall training procedure is smooth, with only minor local fluctuations.

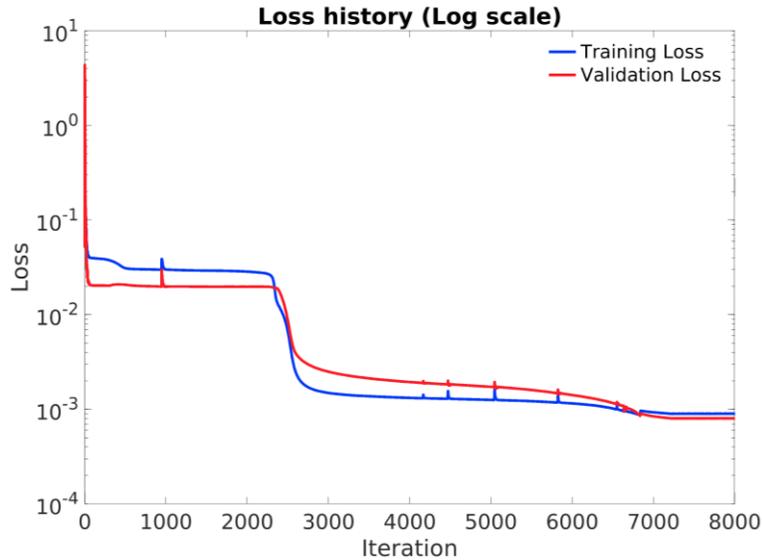

Fig. 11. Loss history for Neural-Rule Model

The predictions made by the trained Neural flow rule model, compared against the experimental data, are presented in Figures 12 and 13. The black dashed line represents the predictions made by the Neural-Rule model, while the blue solid line represents the experimental results. Thanks to the Neural-Rule formulation based on Hooke's law, a good fit is observed in the linear regime, despite some discrepancies under certain loading conditions. We observed that each specimen, representing distinct experimental conditions, exhibits subtle variations in the tangent of the linear elastic portion of the response. These differences could potentially arise from variations in the manufacturing process or minor disparities in the testing environment, especially in load line compliance of the test system. As a result, minor discrepancies persist in the models in the linear elastic domain. While this model approximates the average elastic modulus, its response seems to align more closely with the Chaboche model, exhibiting smoother behavior in comparison to the black box model. However, both the Chaboche and Neural Flow Rule models struggle to capture the overshoot phenomenon during the initial loading cycle. From an experimental perspective, this overshoot may arise due to initial microstructural realignments or transient material behaviors under sudden load applications. Viewing the Neural Flow Rule model as a fusion of data-driven techniques and physical laws, it inherently possesses less flexibility than purely data-driven models like the black box model. However, an advantage of the Neural Flow Rule model is its potential to circumvent overfitting issues, especially when data is limited.

The Mean Square Error (MSE) for the three distinct models is depicted in Fig. 14. The black-box model consistently achieves the lowest error across most loading scenarios, indicating its superior overall accuracy. The Neural Flow Rule model generally performs better than or at least comparably to the standard Chaboche model, though it does register slightly higher errors in a few instances. These findings underscore the efficacy of both the black-box and Neural Flow Rule models in accurately modeling and predicting creep-fatigue material behavior, suggesting that, for the given cases, machine learning models can offer improved predictions compared to the standard Chaboche model. Beyond prediction accuracy, these cases highlight that data-driven machine learning models can reconstruct credible stress-strain relationships and predict a material's creep-fatigue behavior. This insight is pivotal, suggesting that the Neural ODE model can be expanded to address other material behaviors not well-defined by current physical laws or even to more intricate behaviors, such as those displayed by composite materials.

A notable advantage of the Neural ODE approach is its flexibility in parameter initialization. In the standard Chaboche model, every parameter has a specific physical meaning, and an effective range of initial values often needs to be evaluated based on prior experience or expertise. In contrast, the parameters in Neural ODEs are initialized using standard neural network methods, reducing the dependence on prior knowledge. This convenience can be particularly valuable when dealing with novel materials or behaviors where such prior knowledge is limited or unavailable. Future endeavors could concentrate on enhancing these models, with a particular emphasis on the development of more hybrid models. These models would synergistically integrate both physical laws and machine learning methodologies, thereby boosting the model's predictive power while also enhancing interpretability. This approach could unlock new potential for scientific discovery, facilitating a deeper understanding of complex material behaviors and driving the development of more accurate, efficient predictive tools in the field of materials science.

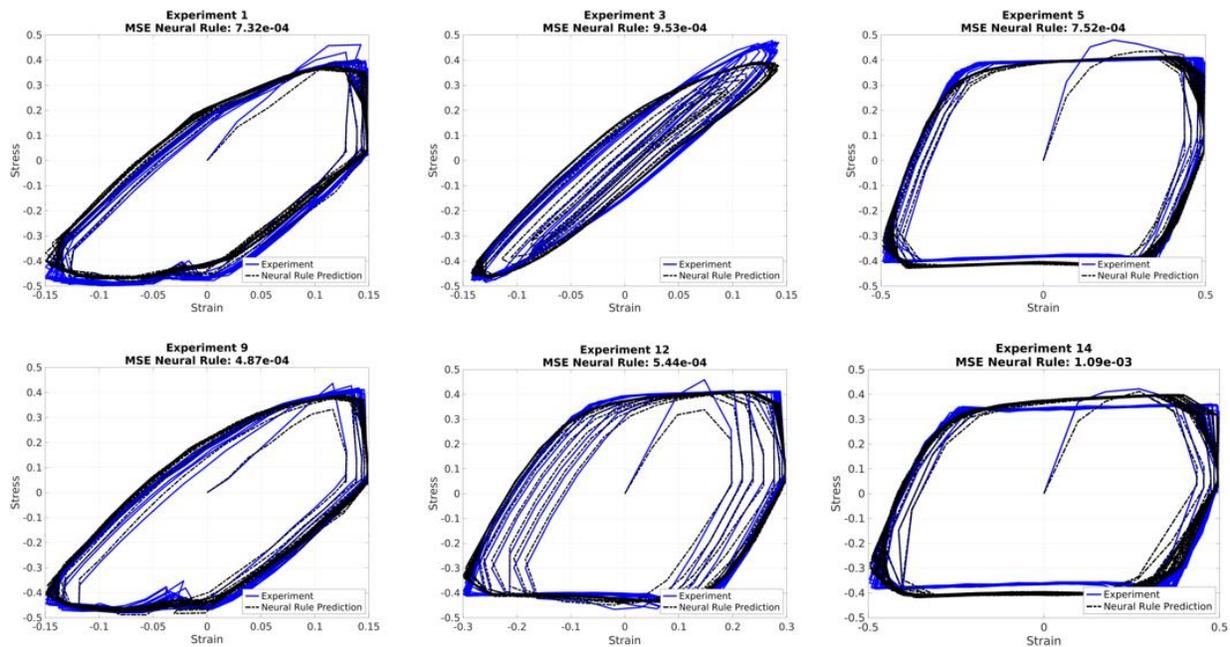

Fig. 12. Comparative Analysis of Fitting Results between the Chaboche Model and the Neural-Rule Model for training dataset

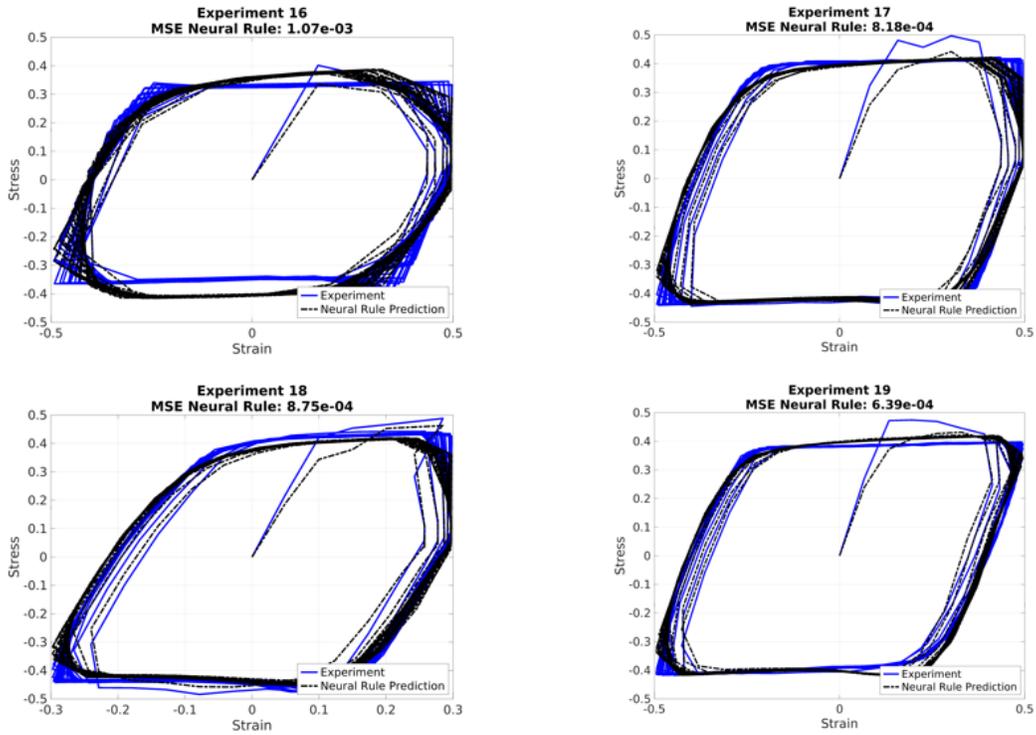

Fig. 13. Neural-Rule Model for validation dataset

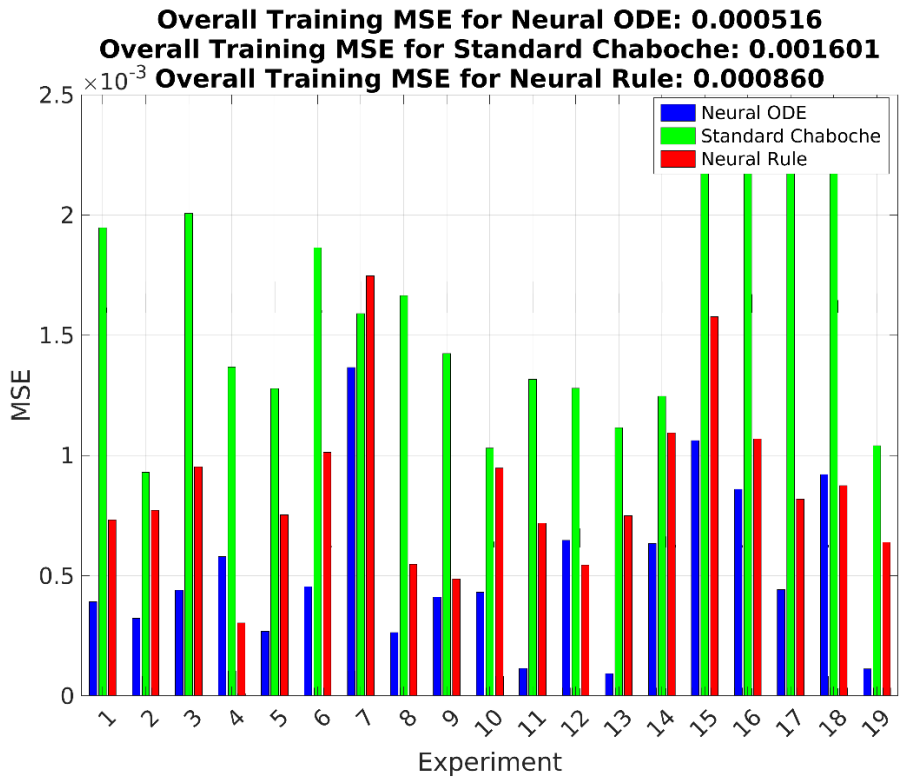

Fig. 14. Comparison of Mean Squared Error between different models

# 5. Interpretable Machine Learning Model

## 5.1 Symbolic Regression for Neural Ordinary Differential Equation

In this part, we utilize symbolic regression to interpret a trained neural ordinary differential equation (ODE) model. Symbolic regression is a powerful method that simultaneously deduces the structure and parameters of a model by exploring a variety of mathematical expressions. This approach contrasts with traditional methods that depend on a predefined model structure, offering a more flexible and insightful analysis. Among the various techniques available for symbolic regression, we employ the Sparse Identification of Nonlinear Dynamical Systems (SINDy) method [31, 32]. SINDy is particularly adept at distilling the essence of complex models. It frames the task of model discovery as a sparse regression challenge, where the aim is to identify the most significant terms in the function $f$ from a comprehensive library of potential functions. This library serves as a reservoir of fixed nonlinear expressions from which the model can be constructed as a linear combination. The strength of SINDy lies in its ability to strike a balance between model accuracy and efficiency. By focusing on sparse solutions, SINDy avoids the pitfall of overfitting, ensuring that the resulting models are both succinct and robust. These models not only perform well but also offer a level of interpretability and generalizability that is crucial for understanding underlying physical processes and for the application in scientific fields where clarity and insight into the model's behavior are paramount.

This approach is intuitive and adaptable, allowing for customization through various sparse regression algorithms or selection of library functions. The SINDy method has gained traction across diverse fields, aiding in model identification for systems ranging from chemical reaction dynamics to nonlinear optics and thermal fluids. In our current study, we employ the SINDy method to elucidate the underlying equations governing creep-fatigue mechanics systems. Essentially, SINDy employs sparse regression to sift through a library of potential features, identifying the optimal combination that best represents the system under investigation. This method not only enhances our understanding of the system's dynamics but also ensures the model's relevance and simplicity by focusing on the most significant features. The core principle of SINDy involves resolving the system given by the equation,

$$\dot{\mathbf{X}} = \Theta(\mathbf{X})\Phi \qquad (16)$$

where we form a library $\Theta$ comprising potential nonlinear functions of the variables contained in $\mathbf{X}$. This library, $\Theta$, might include various functions such as constants and polynomials. $\Phi$ is a sparse vector, indicating which terms are active in the dynamics. For a system dependent on two variables, the function library $\Theta(\mathbf{X})$ could be represented as:

$$\Theta(\mathbf{X}) = [\mathbf{1},\ x_1,\ x_2,\ x_1^2,\ x_2^2,\ x_1 \cdot x_2, sin(x_1), sin(x_2), \cdots] \qquad (17)$$

In this representation, $\Theta(\mathbf{X})$ denotes the suite of candidate functions, with $\mathbf{X}$ encapsulating all the variables of the ODE, including the values of internal variables at different time steps, and $\dot{\mathbf{X}}$ representing the corresponding time derivatives.

Upon successfully training a neural ODE model for a black box system, we can readily extract $\mathbf{X}$ and its time derivative $\dot{\mathbf{X}}$ using the trained neural networks. Following this, we can implement symbolic regression with the SINDy algorithm to perform sparse regression and derive the mathematical expression of the dynamic system. The resulting mathematical equations constitute an interpretable machine learning model, which can greatly enhance our understanding of the system's behavior. In our case, we choose polynomial feature libraries, capping the polynomial order at 3 for our model. To perform sparse regression, we utilize the sequentially threshold least squares method. Selecting an appropriate threshold, $\lambda$, for sparse regression is critical; a smaller $\lambda$ allows for more terms in the model, potentially capturing the system's nuances but risking overfitting. Conversely, a larger $\lambda$ simplifies the model with fewer terms, which may benefit generalization at the expense of precision. Essentially, $\lambda$ serves as a tuning parameter to manage the sparsity of the system. A common strategy for choosing $\lambda$ is to vary it from 0 to 1, evaluating against test data to determine the value that yields the best generalization. For our reconstruction of the black box model, we settle on a $\lambda$ value of 0.05. The fitting outcomes of the interpretable SINDy model, as compared with the experimental data, are presented in Figure 15. For the sake of brevity in the paper, we randomly selected six datasets from the entire experimental collection for comparison. In the Fig. 15, the blue solid line represents the SINDy model's predictions, while the black dashed line corresponds to the experimental data. It is evident that the fitting results of the model closely align with the experimental data.

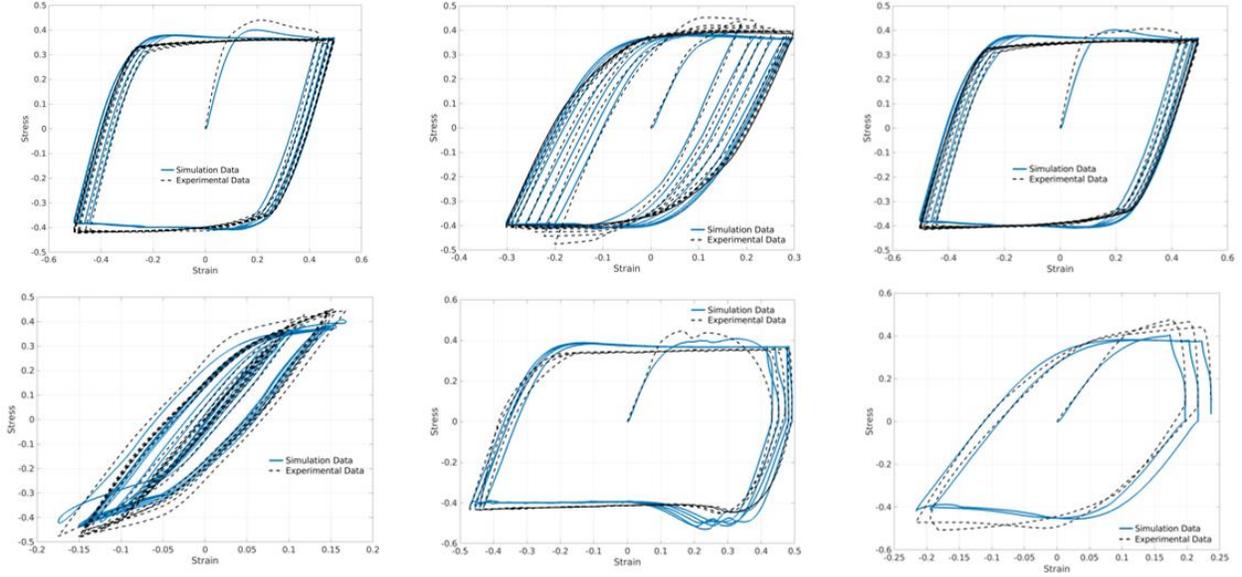

Fig. 15: Comparison of the Interpretable Model with Experimental Data

The interpretable system obtained by the SINDy algorithm can be written as follows:

$$\begin{aligned} \dot{x}_0 &= f_0(x_0, x_1, x_2, x_3) \\ \dot{x}_1 &= f_1(x_0, x_1, x_2, x_3) \\ \dot{x}_2 &= f_2(x_0, x_1, x_2, x_3) \end{aligned} \quad (18)$$

In the above system, $x_3$ represents the external force input; specifically, for our problem, $x_3$ denotes the strain rate. $x_0$ is the observable variable, corresponding to stress, while $x_0$ and $x_1$ serve as internal variables. The detailed mathematical formulation, derived using the SINDy algorithm, is presented as follows:

$$\begin{aligned}
\frac{dx_0}{dt} =& -0.826x_0 - 6.663x_1 + 3.812x_2 - 4.935x_3 + 1.315x_0^2 - 0.738x_0x_1 - 1.504x_0x_2 \\
& + 9.679x_0x_3 - 18.823x_1^2 + 25.164x_1x_2 - 3.155x_1x_3 - 7.780x_2^2 - 0.417x_2x_3 + 32.287x_3^2 \\
& + 0.326x_0^3 - 1.134x_0^2x_1 + 0.160x_0^2x_3 - 7.212x_0x_1^2 + 12.094x_0x_1x_2 + 0.628x_0x_1x_3 \\
& - 2.718x_0x_2^2 - 1.840x_0x_3^2 - 12.730x_1^3 + 35.381x_1^2x_2 - 9.995x_1^2x_3 - 28.937x_1x_2^2 \\
& + 3.144x_1x_2x_3 + 9.813x_1x_3^2 + 6.427x_2^3 - 34.906x_2x_3^2 \\
\frac{dx_1}{dt} =& -2.276x_0 - 18.321x_1 + 10.593x_2 - 27.472x_3 + 0.910x_0^2 - 5.562x_0x_1 + 2.604x_0x_2 \\
& + 1.882x_0x_3 - 40.112x_1^2 + 46.329x_1x_2 - 48.622x_1x_3 - 15.780x_2^2 + 27.881x_2x_3 + 0.226x_0^3 \\
& - 0.518x_0^2x_1 + 5.284x_0^2x_3 - 11.207x_0x_1^2 + 25.825x_0x_1x_2 + 0.368x_0x_1x_3 - 7.757x_0x_2^2 \\
& + 0.594x_0x_2x_3 + 8.167x_0x_3^2 - 5.468x_1^3 + 73.796x_1^2x_2 - 66.752x_1x_2^2 + 91.437x_1x_2x_3 \\
& + 15.256x_2^3 - 34.031x_2^2x_3 \\
\frac{dx_2}{dt} =& 1.732x_0 + 6.453x_1 - 5.013x_2 + 12.238x_3 - 0.246x_0^2 - 9.091x_0x_1 + 4.768x_0x_2 \\
& - 11.866x_0x_3 + 5.449x_1^2 - 36.768x_1x_3 \\
& + 5.449x_1^2 - 36.768x_1x_3 + 26.815x_2x_3 - 34.833x_3^2 - 0.394x_0^3 - 4.118x_0^2x_1 + 1.702x_0^2x_3 \\
& - 18.512x_0x_1^2 + 11.943x_0x_1x_2 + 12.567x_0x_1x_3 + 0.641x_0x_2^2 - 20.345x_0x_2x_3 + 100.083x_0x_3^2 \\
& - 24.888x_1^2x_2 + 48.608x_1^2x_3 + 22.943x_1x_2^2 - 105.871x_1x_2x_3 + 344.996x_1x_3^2 - 7.116x_2^3 \\
& + 45.029x_2^2x_3 - 258.343x_2x_3^2 + 397.446x_3^3.
\end{aligned}$$

The physical system described above was approximated using up to third-order polynomials with the SINDy algorithm. To evaluate the performance of various models, we compared the overall Mean Squared Error (MSE) values across all experiments for different models, ranging from the standard Chaboche model to the black box and neural rule models. Notably, the black box and neural rule models exhibited lower MSE values. However, the MSE of the interpretable model was comparable to that of the standard Chaboche model, indicating a similar level of accuracy. The interpretable model was derived from a black box model, with the assumption of having no knowledge about the physical system. This model was constructed purely from data, using a limited dataset. Despite this, the purely data-driven approach achieved accuracy close to the Chaboche model, an empirical model. This underscores the potential of data-driven methods to uncover complex physical relationships and produce interpretable models with accuracy comparable to that of traditional empirical models.

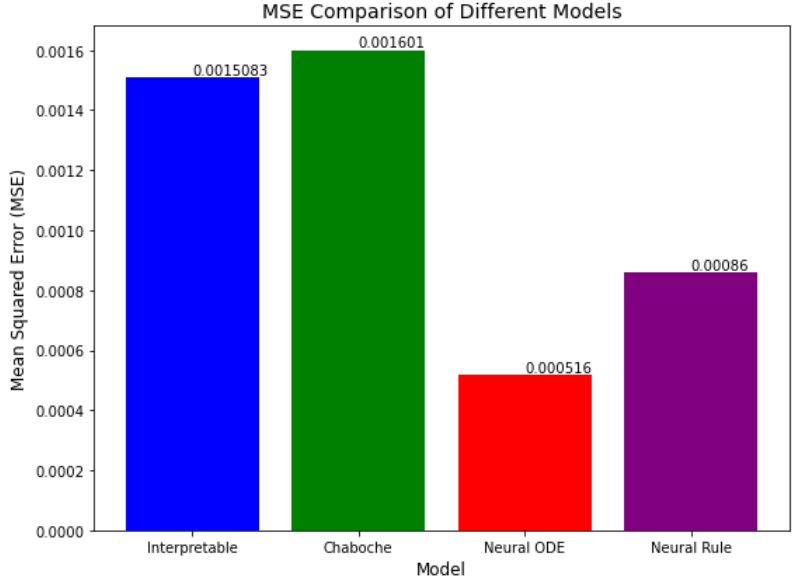

Fig. 16: Mean Squared Error (MSE) Comparison Across Different Models

To further analyze the interpretable model using the state-space method in control theory, we can examine some system stability properties. To linearize a nonlinear system around a specific operating point, we employ the Jacobian matrix to derive the linearized system matrix **A**. This linearization is conducted around a particular operating point, typically represented as $(x^*, u^*)$, where $x^*$ is the state around which we are linearizing, and $u^*$ is the corresponding input at that point. Given the system as depicted in Eq. (14), the linearized system can be expressed as follows:

$$\dot{x} = Ax + Bu \qquad (19)$$

where **A** is the Jacobian matrix of the partial derivatives of the function ff with respect to the states, evaluated at the operating point, and **B** is the Jacobian matrix of the partial derivatives of the function $f$ with respect to the inputs, also evaluated at the operating point. In this case, we are considering the initial state of the mechanical system, which corresponds to [0,0,0] for $x$ and u=0; the Jacobian matrices for such a situation are:

$$A = \begin{bmatrix} -0.826 & -6.663 & 3.812 \\ -2.276 & -18.321 & 10.593 \\ 1.732 & 6.453 & -5.013 \end{bmatrix} \qquad (20)$$

To analyze the stability of matrix **A**, we need to compute the real parts of the eigenvalues of matrix **A**. The real parts of the eigenvalues of matrix **A** are $[-23.259, -0.0321, -0.869]$, all of which are negative. This indicates that the initial state of the system is stable at this operating point. It implies that with a small perturbation at the initial point and a small input $u$, the material can return to its original state, aligning with the fundamental material laws for a linear elastic response. Further analysis can be conducted on this system using the state-space method; however, this would require additional investigation and discussion in future research and is not the primary focus of this paper.

### 5.2 Advantages of Model Interpretability

The polynomial SINDy model provides additional interpretability compared to the neural models. For example, the polynomial for $x_0$ explicitly describes the dependence of the stress rate on the stress, strain, and (arbitrary) internal state. This model also had the advantage of being easily tabulatable: we can compactly provide a written description of the model. The neural models have thousands of parameters and a complex functional formal, which makes it difficult to compactly describe the model.

# 6. Conclusion

The primary aim of this paper is to introduce the use of Neural Ordinary Differential Equation (Neural ODE) models for learning and understanding creep-fatigue material behavior. Previous research primarily focused on material hyperelasticity and low temperature plasticity modeling, while studies concerning creep-fatigue material behavior have been relatively infrequent. Therefore, the application of machine learning models for creep-fatigue characterization represents a critical advancement in this field.

Neural ODE models offer several benefits for modeling physical phenomena. They provide a continuous representation of system dynamics, ensuring better alignment with physical laws and allowing more intuitive interpretations. While machine learning methods often demand extensive training data, Neural ODE models excel in modeling complex material behaviors even with limited data. This advantage is crucial given the challenges in acquiring large experimental datasets, which can contain noise and system errors. Most importantly, neural ODE models can be easily transferred to model material behavior in structural models of full components, for example finite element simulations. Neural ODEs are not different, mathematically, from conventional material models.

The generalization capability of data-driven models is essential for ensuring reliable performance in scenarios beyond those explicitly represented in the training data. In this study, generalization performance is evaluated using the available experimental data, which spans a range of conditions within practical constraints. Due to the limited availability of experimental data and the high cost of obtaining additional samples, out-of-sample testing is not feasible for this paper. Nonetheless, the model's predictive performance across the existing data provides insights into its generalization capabilities within similar conditions.

Compared to standard empirical models (such as the Chaboche model), the Neural ODE approach offers several advantages:

- It can be generalized to a wide category of materials, as it learns the material response directly from the underlying test data.
- While physically-based models have the advantage of clear interpretability due to their inherent physical meanings, developing such models typically requires extensive expert knowledge and experience. A promising future direction could involve making the Neural ODE model more interpretable, or exploring more intricate hybrid models that better integrate physical laws.
- The black box neural model is able to describe complex details of material response, such as the overstress behavior in the experiments, especially for the first cycle.
- The black box neural model is more accurate in general, compared to the standard Chaboche model, for this data set.
- The interpretable model obtained through symbolic regression can achieve accuracy comparable to the standard Chaboche model. Compared to other time series ML models, the neural ODE model is more explainable, offering a kind of interpretability that provides a deeper understanding of the physical system.

Looking ahead, we anticipate extending these machine learning models to handle more complex material behaviors. The development of more advanced hybrid models that integrate the strengths of physical laws and machine learning could further enhance accuracy, thus advancing our capabilities in predictive materials science.

## Acknowledgement

The research was sponsored by the U.S. Department of Energy under Contract No. DE-AC02-06CH11357 with Argonne National Laboratory, managed and operated by UChicago Argonne LLC. Funding was provided by the U.S. Department of Energy, Office of Nuclear Energy.